\documentclass[11pt]{article}
\usepackage{graphics,epsfig,color}
\usepackage[]{graphicx}
\usepackage{latexsym}
\usepackage{amsmath, amsthm, amsfonts, amssymb}
\usepackage{verbatim}

\newtheorem{theorem}{Theorem}

\begin{document}

%%%%%%Decio's commands%%%%%%%%%%
\newcommand{\ext}{=_E}
\newcommand{\Q}{$\mathfrak{Q}$}
\newcommand{\ita}{\textit}
\newcommand{\lra}{\leftrightarrow}
%%%%%%%%%%%%%%%%%%%%%

\newtheorem{theo}{Theorem}[section]
\newtheorem{definition}[theo]{Definition}
\newtheorem{lem}[theo]{Lemma}
\newtheorem{prop}[theo]{Proposition}
\newtheorem{coro}[theo]{Corollary}
\newtheorem{exam}[theo]{Example}
\newtheorem{rema}[theo]{Remark}
\newtheorem{example}[theo]{Example}
\newtheorem{principle}[theo]{Principle}
\newcommand{\ninv}{\mathord{\sim}} %involutive negation
\newtheorem{axiom}[theo]{Axiom}
\numberwithin{equation}{subsection}

\title{Quantum Logical Structures For Identical Particles}

\author{{\sc Federico Holik}$^1$ \ {\sc ,} \ {\sc Ignacio Gomez}$^2$ {\sc and} \ {\sc D\'{e}cio Krause}$^3$}

\maketitle

\begin{center}

\begin{small}
1- Instituto de F\'{\i}sica La Plata, CCT-CONICET, and Departamento
de F\'{\i}sica, Facultad de Ciencias Exactas, Universidad Nacional
de La Plata - $115$ y $49$, C.C.~$67$, $1900$ La Plata, Argentina \\

2- Departamento de Matem\'{a}tica - Facultad de Ciencias Exactas y
Naturales\\ Universidad de Buenos Aires - Pabell\'{o}n I, Ciudad
Universitaria \\ Instituto de F\'{\i}sica de Rosario (IFIR-CONICET), Rosario, Argentina\\

3-Department of Philosophy\\ Federal University of Santa Catarina,\\
Florian\'opolis, SC - Brazil.
\end{small}
\end{center}

\vspace{1cm}

\begin{abstract}
\noindent In this work we discuss logical structures related to
indistinguishable particles. Most of the framework used to develop
these structures was presented in
\cite{extendedql,Holik-Massri-Ciancaglini-2010} and in
\cite{French-Kra-2006,Dom-Hol-2007,Dom-Hol-Kra-2008,Dom-Hol-Kniz-Kra-2009}.
We use these structures and constructions to discuss possible
ontologies for identical particles. In other words, we use these
structures in order to characterize the logical structure of quantum
systems for the case of indistinguishable particles, and draw
possible philosophical implications. We also review some proposals
available in the literature which may be considered within the
framework of the quantum logical tradition regarding the problem of
indistinguishability. Besides these discussions and constructions,
we advance novel technical results, namely, a lattice theoretical
structure for identical particles for the finite dimensional case.
This kind of approach was not present in the scarcely literature of
quantum logic and indistinguishable particles.
\end{abstract}
\bigskip
\noindent

\begin{small}
\centerline{\em Key words: quantum logic-convex
sets-indistinguishable particles}
\end{small}

\bibliography{pom}

\section{Introduction}

Among the foundational problems of quantum mechanics ($QM$), the
discussion about identical particles (so denominated for example by
van Fraassen \cite[chaps.11 and 12]{van91}) plays a central role.
The case of ``indistinguishable" particles is treated separately in
almost all introductory books on $QM$ \cite{Ballentine,Sakurai}, and
many works on physics
\cite{SchroWhatIs?,MessiahGreenberg-1964,Girardeau,IdenticalAmico,IdenticalDowling,IdenticalEntGirardi,IdenticalPlastino}
and philosophical debate \cite{sch52,d'esp,French-Kra-2006,Heinz
Post,Why quasisets} has been dedicated to this problem.

From the interpretational point of view, one the of the most
important open tasks is the right characterization of the word
`identical' as used in this context. On the one hand, the axioms of
QM, as standardly formulated, are based on classical logic and
mathematics: that mathematics that can be build in a fragment of the
Zermelo-Fraenkel set theory with the axiom of choice (ZFC)
\cite{Manin-SetTheory,Halmos,Kunen}. Thus, it involves the
`classical' theory of identity, which says that there are no
indiscernible objects: indiscernible, or indistinguishable, entities
must be \textit{the very same object}, and that's all. But on the
other hand, the theory does not seem to provide the means to
distinguish (or label) quanta in most cases. On the contrary, it
seems to point in the direction that quanta cannot be discerned at
all. Thus, at the interpretational level, a lot of authors inclined
themselves for a point of view in which \ita{quanta} \emph{cannot}
be considered individuals
\cite{sch52,SchroWhatIs?,French-Kra-2006,frekra11,kra10,Why
quasisets,Heinz Post}. This position is usually called \emph{the
received view}. The question either elementary particles may be
indiscernible by `physical properties' only (but retain
individuality), or if they can be absolutely indiscernible, being
entities that even in God's mind cannot be discerned, either by
physical or logical tools, is not a settled topic (see for example
\cite{MullerSaubders-2008} and \cite{MullerSeevinck-2009} and
\cite{French-Kra-2006} for a complete discussion). We will return to
this problem in this paper and use logical tools to shed light into
it.

Many authors claim that the alleged indiscernibility concerns only
with `physical properties', and usually state that even if quanta
cannot be discerned by physical means, they still retain their
individuality as some kind of ``primitive thisness". Thus, as we
shall see, this alternative opens the door for some kind of hidden
variables, or parameters (in this case, as a form of ``hidden
identity"), which would exist at least in the logical domain --logic
here encompasses mathematics-- and which hide themselves when we try
to manipulate them experimentally (as happens with ``hidden
variables" in Bohm's interpretation). No one knows the implications
of this kind of assumption, as for instance, if there is some kind
of non-go theorem speaking of `logical' properties, and confusion
usually appears in crucial questions, such as testing experimentally
if quanta are individuals or not. In order to settle such questions
the problem must be properly formulated and some questions
clarified. Logic may be helpful for this task, and we will explore
this possibility throughout the paper.

If the received view is accepted, there is in fact a foundational
problem of logical nature when the interpretation -which presupposes
non individuality- is contrasted with the standard axiomatic
formulation of the theory, which as mentioned above, presupposes the
classical theory of identity. This situation gave rise to different
kinds of criticisms, as the discussion about the ``surplus
structure" in \cite{redtel91,redtel92}, the demand of a more direct
formulation (avoiding the symmetrization postulate) as in \cite{Why
quasisets,Heinz Post}, or the simpler exploitation of the
contradiction in order to support the position that quanta are
individuals (at least in a weak form), as in
\cite{MullerSeevinck-2009}. As we shall see in this work, it is
possible to put things clearer by stating the problem properly by
logical means, in the case, by changing the underlying logic.

Another important foundational problem in $QM$ is the one of
compound systems (of which the indistinguishability problem may be
considered as a particular case). In particular, the notion of
entanglement \cite{bengtssonyczkowski2006}, which was considered by
Schr\"{o}dinger as ``{\bf the} characteristic trait of quantum
mechanics"\cite{Schro-35,Schro-36,EPR}. Entanglement has to do with
the properties of quantum systems when they interact, when they are
gathered together, and studies which concentrate on entanglement of
identical particles have been developed relatively recently. Many
questions remain open and, as is well known, correlations originated
in entanglement are very different than those originated in
statistics (exchange correlations)
\cite{IdenticalEntGirardi,IdenticalPlastino,IdenticalAmico,IdenticalDowling}.
The singular features which appear when aggregates of systems of
identical particles are studied makes the subject to have its own
particular problems. This has to be taken into account when
structural properties of quanta aggregates are studied.

\emph{In this paper, we shall outline a discussion on certain topics
involving indiscernible particles within the scope of different
mathematical structures erected from motivations taken from quantum
theory.} One of them has to do with \emph{quantum logics} ($QL$) and
compound quantum systems of identical particles. The standard
quantum logical approach to $QM$
\cite{BvN,dallachiaragiuntinilibro,jauch,piron,belcas81,dvupulmlibro,HandbookofQL},
uses the lattice of projections of the Hilbert space of the system
as the lattice of propositions (see Section \ref{s:introduction to
QL}). This approach has been useful for the study of structural
properties of quantum systems by the characterization of their
operational lattices, and to clarify differences with other theories
(such as classical mechanics). Recently, an alternative proposal has
been developed
\cite{extendedql,Holik-Massri-Ciancaglini-2010,Holik-Massri-Plastino-Zuberman}
in order to solve some problems which appear in the study of
compound quantum systems (see for example \cite{aertsparadox}).
After reviewing the standard formulation of the formalism of
indistinguishable particles and posing its problems in Section
\ref{s:IdenticalParticles}, we will adapt the constructions
presented in
\cite{extendedql,Holik-Massri-Ciancaglini-2010,Holik-Massri-Plastino-Zuberman}
to the indistinguishable particle case in section
\ref{s:introduction to QL}. This construction is particularly
suitable for an extension of the $QL$ approach to the case of
indistinguishable particles, which is difficult to accommodate in
the traditional approach, and was not explored in the literature
(though see \cite{Grigore}). It provides a new formal framework for
the study of compound quantum systems in the indistinguishable case
and its entanglement properties, as discussed in section
\ref{s:Convex lattice}.

The other formal structure studied in this work is the theory of
quasi-sets \cite{Why quasisets}, which is based on a non-classical
logic, namely, a non-reflexive logic, and has to do with the problem
of the identity of indistinguishable particles\footnote{We call
`non-reflexive' those logics which deviate from classical logic in
what respects the theory of identity, in particular, by questioning
the principle of identity in some formulation. In our case, we
question the principle in the form $\forall x (x=x)$ (also called
the reflexive law of identity) since we assume that there are
entities to which the standard notion of identity (described by the
theory governing the symbol `=') does not hold. This assumption is
based on some of Schr\"odinger's opinions --see \cite{sch52} and
\cite{French-Kra-2006} for a detailed history and context.}. This
will be done in section \ref{s:Quasi-sets}. In this section we will
review the main characteristics of quasi-set theory and discuss its
implications. We will also review how it can be used for an
alternative formulation of the Fock-space formalism
\cite{Dom-Hol-Kra-2008,Dom-Hol-Kniz-Kra-2009} discussing the
implication of such a construction for he interpretation of $QM$.
Next, in Section \ref{s:WhatIsToBeDone?}, we pose the problem of
identical particles in a new form under the light of the logical
structures presented in this work.

Finally, we will present our conclusions in Section
\ref{s:OntologicalImplications}, where we will try to condense some
ontological implications of the discussions posed in Sections
\ref{s:introduction to QL}, \ref{s:Quasi-sets} and
\ref{s:WhatIsToBeDone?}.

\section{The problem of identical
particles}\label{s:IdenticalParticles}

We shall sketch here a small introduction to the standard
formulation of the problem of identical particles. We will emphasize
the usual mathematical trick that is used to achieve
indistinguishability, namely, permutational symmetry. The
clarification of that trick opens the door --from the foundational
point of view-- to the idea that a different mathematical formalism
would be in order. We will explore the (old) idea that physics may
suggest new logical schemas (see for example \cite{Putnam}).

\subsection{States And Compound Quantum Systems}

In the standard quantum mechanical formalism, for any system $S$, a
Hilbert space $\mathcal{H}$ is assigned and observables are
represented by self adjoint compact operators. Let
$\mathcal{B}(\mathcal{H})$ denote the set of bounded operators on a
suitable Hilbert space $\mathcal{H}$, while the set of bounded self
adjoint operators is denoted by $\mathcal{A}$.
$\mathcal{B}(\mathcal{H})$ is a well known example of a von Neumann
algebra \cite{mikloredeilibro}.

States will be represented (for either single or compound systems)
by the set of positive trace class and self adjoint operators of
trace $1$,

\begin{equation}\label{e:ConvexSet}
\mathcal{C}=\{\rho\in\mathcal{A}\,|\,\mbox{tr}(\rho)=1 \,\,
\mbox{and}\,\,\rho\geq 0\},
\end{equation}

\noindent where the operators $\rho$ are called \textit{density
operators}. They represent the more general available states, and
for any observable represented by  an hermitian operator $A$, they
assign a real number (which is interpreted as the mean value of the
observable) according to the rule

\begin{equation}\label{e:BornRule}
\mbox{tr}(\rho A)=\langle A\rangle
\end{equation}

\noindent If $P$ is a projection operator (i.e., it satisfies
$P^2=P$) intended to represent an elementary test or event
\cite{BvN,piron}, then the real number $\mbox{tr}(\rho P)$ is
interpreted as the probability of obtaining the event $P$ given the
system in state $\rho$. This is no other thing than the Born's rule.

There is a special subclass of states, called pure states, which are
the density operators satisfying $\rho^2=\rho$. They have a
representation as normalized vectors in $|\psi\rangle\in\mathcal{H}$
in the form

\begin{equation}
\rho_{pure}=|\psi\rangle\langle\psi|
\end{equation}

\noindent States which are not pure are called \emph{mixed states}.
For pure states, we have the superposition principle: any normalized
linear combination of states will yield a new state. In formulae, if
$|\psi\rangle$ and $|\varphi\rangle$ are normalized vectors
representing pure states, and if $\alpha$ and $\beta$ are complex
numbers satisfying $|\alpha|^2+|\beta|^2=1$, then

\begin{equation}
|\phi\rangle=\alpha|\psi\rangle+\beta|\varphi\rangle
\end{equation}

\noindent will also be a state.

For a compound quantum system formed by two subsystems represented
by Hilbert spaces $\mathcal{H}_{1}$ and $\mathcal{H}_{2}$, we assign
a Hilbert space $\mathcal{H}=\mathcal{H}_{1}\otimes\mathcal{H}_{2}$,
where ``$\otimes$" denotes the \emph{tensor product}. Observables
are represented by Hermitian operators in
$\mathcal{B}(\mathcal{H}_{1}\otimes\mathcal{H}_{2})$. Let
$\{|\varphi_{i}^{(1)}\rangle\}$ and $\{|\varphi_{i}^{(2)}\rangle\}$
be orthonormal basis of $\mathcal{H}_{1}$ and $\mathcal{H}_{1}$
respectively. The set
$\{|\varphi_{i}^{(1)}\rangle\otimes|\varphi_{j}^{(2)}\rangle\}$
forms an orthonormal basis for
$\mathcal{H}_{1}\otimes\mathcal{H}_{2}$. Then, a pure state of the
composite system can be written as
$|\psi\rangle=\sum_{i,j}\alpha_{ij}|\varphi_{i}^{(1)}\rangle\otimes|\varphi_{j}^{(2)}\rangle$.
Given a state $\rho$ of the composite system, partial states
$\rho_1$ and $\rho_2$ can be defined for the subsystems. The
relation between $\rho$, $\rho_{1}$ and $\rho_{2}$ is given by:

\begin{equation}
\rho_{1}=tr_{2}(\rho) \ \ \ \ \rho_{2}=tr_{1}(\rho)
\end{equation}

\noindent where $tr_{i}$ stands for the partial trace over the $i$
degrees of freedom. A density matrix of the composite system $\rho$
is said to be a convex combination of product states, if there
exists $\{p_{i}\}$ and states $\{\rho_{1}^{i}\}$ and
$\{\rho_{2}^{i}\}$ such that

\begin{equation}
\rho=\sum_{i=1}^{N}p_i\rho_{1}^{i}\otimes\rho_{2}^{i}
\end{equation}

\noindent If a state of the composite system can be written as a
convex combination of product states (or approximated by a sequence
of them), then it is said to be \emph{separable}. If not, it is said
to be entangled \cite{bengtssonyczkowski2006}.

\subsection{Indistinguishable particles}

If particles are identical -in the sense of sharing all their
intrinsical properties (for example, a collection in which all
particles are electrons)-, we must add the condition that all pure
states should be symmetrized. This is the content of the
symmetrization postulate \cite{French-Kra-2006}, and this means that
all states must have a definite symmetry with respect to the action
of the permutation operator. For example, if $|\varphi\rangle\in
\mathcal{H}_{1}$, $|\psi\rangle\in \mathcal{H}_{2}$ and
$|\varphi\rangle\neq|\psi\rangle$, then the corresponding
symmetrized states are

\begin{equation}\label{e:Symmetrized}
|\psi\rangle=\frac{1}{\sqrt{2}}(|\varphi\rangle\otimes|\phi\rangle\pm|\phi\rangle\otimes|\varphi\rangle)
\end{equation}

\noindent where the ``$+$" sign stands for bosons and the ``$-$"
sign for fermions. If $\{|\varphi_{i}\rangle\}_{i\in I}$ and
$\{|\phi_{j}\rangle\}_{j\in I}$ are basis of $\mathcal{H}_{1}$ and
$\mathcal{H}_{2}$ respectively, then
$\{|\varphi_{i}\rangle\otimes|\phi_{j}\rangle\}_{<i,j>\in I\times
I}$ is a basis of $\mathcal{H}$. Then, a permutation operator

\begin{eqnarray}
&P_{12}:\mathcal{H}\longrightarrow\mathcal{H}\nonumber\\
&P_{12}|\varphi_{i}\rangle\otimes|\phi_{j}\rangle\rightarrow|\phi_{j}\rangle\otimes|\varphi_{i}\rangle&
\end{eqnarray}

\noindent can be defined, because it is defined on each element of
this basis (and extended linearly in a trivial way). It can be shown
that it is independent of the chosen basis. As
$P_{12}^{2}=\mathbf{1}$, its eigenvalues are $1$ and $-1$ for Bosons
and Fermions respectively. Then, this operator selects two special
subspaces of $\mathcal{H}$ according to its eigenvalues

\begin{subequations}
\begin{equation}
\mathcal{H}^{+}=\{|\psi\rangle\in
\mathcal{H}\,|\,P_{12}|\psi\rangle=|\psi\rangle\}
\end{equation}
\begin{equation}
\mathcal{H}^{-}=\{|\psi\rangle\in
\mathcal{H}\,|\,P_{12}|\psi\rangle=-|\psi\rangle\}
\end{equation}
\end{subequations}

\noindent which will represent the possible (pure) states of the
system when indistinguishable particles are involved. All physical
states of indistinguishable particles must obey these symmetry
conditions. This is an empirical statement, and up to now, no other
symmetries where found (see \cite{Girardeau,MessiahGreenberg-1964}
for a discussion of this statement). So, the theory opens the door
to ``para-statistics", but it seems that none of them where found to
have correspondents in nature, and we will not treat this case here.

\subsection{How do the symmetrization postulate works and its open
problems}\label{s:PosingTheProblem}

It is important to make here the crucial observation of how the
scheme of the symmetrization postulate works. First, particles are
\emph{labeled} by assigning them normalized vectors $|\psi\rangle$
and $|\phi\rangle$ in their corresponding spaces $\mathcal{H}_{1}$
and $\mathcal{H}_{2}$. If the state of the compound system would be
simply $|\psi\rangle\otimes|\phi\rangle$, then, particles could be
distinguished by special observables, i.e., the theory would allow
for an asymmetry between both systems. But things are not so, and
the symmetrization postulate must erase any obserbable
characteristic which allows us to identify the particles. Thus, the
state must be symmetrized as in \ref{e:Symmetrized}. This is how the
symmetrization postulate works: by first imposing a label as a mean
to \emph{individuate} each particle, and then erasing it. It is not
difficult to realize that this trick is unavoidable in the standard
formulation of QM, given that its axiomatic is formulated using
standard (Zermelo-Frenkel) mathematics, based on classical theory of
identity. Thus, symmetrization postulate \emph{hides} particle
identities, living the door open to an interpretation based on
non-individuals. This is one of the reasons why many authors support
the received view.

But the received view has been criticized in many ways. One of them
has to do with contradiction between the method of introducing the
adequate symmetries and the interpretation itself. On the one hand,
Schr\"{o}dinger and others tell us that we must give up any intent
to provide individuality for elementary particles, and this position
has a considerable agreement with experience. But it is also true
that particles are labeled in the \emph{symmetrization mechanism},
and thus, they seem to posses some form of individuality. How to
reconcile these views?

The problem was discussed and a possible solution was proposed in
\cite{redtel91,redtel92}. The authors, characterized the
non-symmetrical parts of the Hilbert space, discarded by the
symmetrization postulate, as \emph{surplus structure}, i.e., a
mathematical structure which plays no role in the final formulation
of the theory and its predicted experience. Indeed, this is true.
Next, they show arguments in favor of the Fock-space formulation of
quantum mechanics. As is well known, it is possible to use the
Fock-space formalism as an alternative approach to $QM$
\cite{Robertson}. The criticism raised against this solution,
asserts that the Fock-space mechanism also appeals to particle
labeling \cite{French-Kra-2006}, and so, it has a similar problem to
that of the usual symmetrization postulate formulation.

From a different point of view, in \cite{MullerSaubders-2008} and
\cite{MullerSeevinck-2009}, it is argued that fermions \emph{are}
individuals, at least in a weak form, because antisymmetry grants
that they have opposite properties: think in two electrons in the
ground state of a Helium atom, they have opposite spin, and thus,
there is a property which one has and not the other. In a later
work, they ``show" that particles are individuals just because
stating the axioms of the standard formulation of $QM$ it can be
shown that particles can be identified. As we shall discuss in
detail in Section \ref{s:Quasi-sets}, this reduces to the fact that
$QM$ is axiomatized using $ZF$ mathematics, and we will argue that
the conclusion drawn by the authors is premature and the problem and
its alternative solutions are not properly posed.

As we shall see later on this work, there is still a possibility for
a reconciliation between the received view and a valid reformulation
of $QM$, but one which encompasses a radical change in the logical
framework of the theory. This will be done in Sections
\ref{s:Quasi-sets} and \ref{s:OntologicalImplications}, after
clarifying the problem, i.e., planting it in a clear logical and
ontological form.

One question is still at stack. Is the received view really
desirable? Or unavoidable? The assumption of the individuality of
quanta seem to play no role in any experience or at least, it can be
removed and experiments can be successfully explained. On the
contrary, if the assumption of individuality is not taken with
special care and protected by suitable hypotheses, it may lead to
wrong results. Then, in spite of this, why not still postulate a
form of ``hidden individuality", playing no role but satisfying a
particular metaphysical taste? As we shall see, this can be done,
and we will discuss the consequences of this assumption when things
are properly formulated.

\section{A quantum logical formalism for identical particles}\label{s:introduction to QL}

In this section, we introduce the lattice of convex subsets
formalism presented in
\cite{extendedql,Holik-Massri-Ciancaglini-2010,Holik-Massri-Plastino-Zuberman}.
Next, we apply it to the case of identical particles using the
preliminaries introduced above. The traditional approach to quantum
logic uses the bounded projection operators (or equivalently, closed
subspaces) of the Hilbert space as propositions or properties, using
the direct sum as the disjunction, the intersection as the
conjunction, inclusion as the order relation and orthogonal
complement as negation
\cite{BvN,dallachiaragiuntinilibro,HandbookofQL}. Contrary to that,
we will use a lattice formed by convex subsets of the state space.
This lattice is more suitable in order to define maps which relate
states of a compound system to states of the subsystems, and allows
for the introduction of mixed states (something almost unavoidable
when we have for example, bipartite systems of identical particles)
\cite{Holik-Massri-Ciancaglini-2010,Holik-Massri-Plastino-Zuberman}.
It is important to remark that no similar construction can be made
with the usual projection lattice for the identical particles case
(as shall be clear from the discussions below).

\subsection{The Lattice of Convex Subsets}\label{s:Convex lattice}

According to the orthodox quantum logical approach, the propositions
of classical mechanics are the subsets of \emph{the set} of states
(classical phase space) \cite{dallachiaragiuntinilibro}. In
\cite{Holik-Massri-Ciancaglini-2010,Holik-Massri-Plastino-Zuberman}
the \emph{convex} subsets of the \emph{convex} set of states are
considered. Convexity is an key feature of quantum mechanics (see
for example \cite{MielnikGQS}, \cite{MielnikTF} and
\cite{MielnikGQM} for an axiomatization based on convex sets). Let
us begin by considering the set of all convex subsets of
$\mathcal{C}$ (defined in equation (\ref{e:ConvexSet}))

\begin{definition}
$\mathcal{L}_{\mathcal{C}}:=\{C\subseteq\mathcal{C}\,|\, \mbox{C is
a convex subset of} \,\,\,\mathcal{C}\}$
\end{definition}

\noindent In order to give $\mathcal{L}_{\mathcal{C}}$ a lattice
structure, we introduce the following operations:

\begin{definition}\label{definitionlattice}

For all $C,C_1,C_2\in\mathcal{L}_{\mathcal{C}}$

\begin{enumerate}
\item [$(\wedge)$] \;
$C_1\wedge C_2:= C_1\cap C_2$

\item[$(\vee)$] \;
$C_1\vee C_2:=conv(C_1,C_2)$. It is again a convex set, and it is
included in $\mathcal{C}$ (using convexity).

\item[$(\neg)$] \;
$\neg C:=C^{\perp}\cap\mathcal{C}$

\item[$(\longrightarrow)$] \;
$C_1\longrightarrow C_2:= C_1\subseteq C_2$

\end{enumerate}

\end{definition}

\noindent With the operations of definition \ref{definitionlattice},
it is apparent that $(\mathcal{L}_{\mathcal{C}};\longrightarrow)$ is
a poset. If we set $\emptyset=\mathbf{0}$ and
$\mathcal{C}=\mathbf{1}$, then,
$(\mathcal{L}_{\mathcal{C}};\longrightarrow;\mathbf{0};\mathbf{1})$
will be a bounded poset. With the operations defined in
\ref{d:definitionlattice}, $\mathcal{L}_{\mathcal{C}}$ will be a
bounded, atomic and complete lattice.

It is possible to define maps which connect states of the compound
system with states of the subsystems as follows
\cite{Holik-Massri-Ciancaglini-2010}. Given
$C_{1}\subseteq\mathcal{C}_{1}$ and $C_{2}\subseteq\mathcal{C}_{2}$,
define

\begin{equation}
C_1\otimes C_2:=\{\rho_{1}\otimes\rho_{2}\,|\,\rho_{1}\in
C_1,\rho_{2}\in C_2\}
\end{equation}

\noindent Using the above definition, it is possible to define the
map

\begin{eqnarray}
\Lambda:\mathcal{L}_{\mathcal{C}_{1}}\times\mathcal{L}_{\mathcal{C}_{2}}\longrightarrow\mathcal{L}_{\mathcal{C}}\nonumber\\
(C_{1},C_{2})\longmapsto conv(C_1\otimes C_2)
\end{eqnarray}

\noindent If we use partial traces:

\begin{eqnarray}
\mbox{tr}_{i}:\mathcal{C}\longrightarrow \mathcal{C}_{j}\nonumber\\
\rho\longmapsto \mbox{tr}_{i}(\rho)
\end{eqnarray}

\noindent we can also construct the maps

\begin{eqnarray}
\tau_{i}:\mathcal{L}_{\mathcal{C}}\longrightarrow\mathcal{L}_{\mathcal{C}_{i}}\nonumber\\
C\longmapsto \mbox{tr}_{i}( C )
\end{eqnarray}

\noindent which link the elements of $\mathcal{L}_{\mathcal{C}}$
(compound system) to the elements of $\mathcal{L}_{\mathcal{C}_{1}}$
and $\mathcal{L}_{\mathcal{C}_{2}}$ (subsystems).

\noindent Next, define the product map

\begin{eqnarray}\label{e:Tau}
\tau:\mathcal{L}_{\mathcal{C}}\longrightarrow\mathcal{L}_{\mathcal{C}_{1}}\times\mathcal{L}_{\mathcal{C}_{2}}\nonumber\\
C\longrightarrow(\tau_{1}(C),\tau_{2}(C))
\end{eqnarray}

\noindent In
\cite{Holik-Massri-Ciancaglini-2010,Holik-Massri-Plastino-Zuberman}
it is shown that using $\Lambda$ and $\tau$ it is possible to link
states of the compound system to the states of its subsystems (at
the lattice level). This cannot be done in the standard QL
formalism. In the following we will take this feature of
$\mathcal{L}_{\mathcal{C}}$ as an advantage to construct a lattice
for the case of identical particles, in which the use of mixtures is
unavoidable.

\subsection{The Identical Particle Lattice $\mathcal{L}_{\mathcal{C}^{\pm}}$
of Convex Subsets (Bipartite Case)}\label{s:Convex lattice}

Let us now define a lattice for the identical particles case. We
will restrict to the finite dimensional bipartite case for
simplicity. We begin by building the lattice of convex subsets for
symmetrized Hilbert spaces. Taking into account the principle of
indistinguishability, let

\begin{definition}
$\mathcal{C}^{\pm}=\{\rho:\mathcal{H}^{\pm}\longrightarrow
\mathcal{H}^{\pm}\,|\,\,t_r(\rho)=1\,\,,\rho^{\dag}=\rho\,\,\mbox{and}\,\,
\rho \geq 0 \}$
\end{definition}

\noindent and define (in analogy with $\mathcal{L}_{\mathcal{C}}$)

\begin{definition}
$\mathcal{L}_{\mathcal{C}^{\pm}}:=\{C\subseteq\mathcal{C}^{\pm}\,|\,
\mbox{C is a convex subset of} \,\,\,\mathcal{C}^{\pm}\}$
\end{definition}

\noindent $\mathcal{C}^{\pm}$ can be considered as a convex subset
of $\mathcal{C}$, namely $\mathcal{C}^{\pm}\in
\mathcal{L}_{\mathcal{C}}$. This is because any matrix in
$\mathcal{C}^{\pm}$ can be canonically extended to the whole Hilbert
space. In order to provide $\mathcal{L}_{\mathcal{C\pm}}$ with a
lattice structure similar to that of $\mathcal{L}_{\mathcal{C}}$, we
define the following operations:

\begin{definition}\label{d:definitionlattice} For any
$C,C_1,C_2\in\mathcal{L}_{\mathcal{C}^{\pm}},$

\begin{enumerate}

\item[$(\wedge^{\pm})$]
$C_1\wedge^{\pm} C_2:= C_1\cap C_2$

\item[$(\vee^{\pm})$]
$C_1\vee^{\pm} C_2:=conv(C_1,C_2)$

\item[$(\neg^{\pm})$]
$\neg^{\pm} C:=C^{\perp}\cap\mathcal{C^{\pm}}$

\item[$(\longrightarrow^{\pm})$]
$C_1\longrightarrow^{\pm} C_2:= C_1\subseteq C_2$

\end{enumerate}
\end{definition}

\noindent $C^{\perp}$ is the orthogonal complement of $C$ with
respect to the scalar product $\langle
A,B\rangle=\mbox{tr}(A.B^{\dag})$, namely $C^{\perp}=\{\rho\in
\mathbb{C}^{N\times N}\,\,|\,\,\mbox{tr}(\sigma.\rho^{\dag})=0 \,\,
\,\,\forall \sigma\in C\}$. With these operations, it follows that
$(\mathcal{L}_{\mathcal{C}^{\pm}};\longrightarrow^{\pm})$ is a
partially ordered set. And if we take $\emptyset=\mathbf{0}$ and
$\mathbf{1}=\mathcal{C}^{\pm}$, then
$(\mathcal{L}_{\mathcal{C}^{\pm}};\longrightarrow^{\pm};\mathbf{0};\mathbf{1})$
is a bounded partially ordered set. We also notice that, because
$\mathcal{C}^{\pm}\in\mathcal{L}_{\mathcal{C}}$ and since the
operations of $\mathcal{C}^{\pm}$ are inherited from $\mathcal{C}$,
then $\mathcal{L}_{\mathcal{C}^{\pm}}$ is a sublattice of
$\mathcal{L}_{\mathcal{C}}$.

\noindent Let
$\mbox{tr}_i(\mathcal{L}_{\mathcal{C}^{\pm}}):=\{\mbox{tr}_i(C)\,\,|\,\,C\in\mathcal{L}_{\mathcal{C}^{\pm}}\}\subseteq\mathcal{L}_{\mathcal{C}_j}$
($i\neq j$). It is possible to define the canonical projections
$\tau_i$ and $\tau$ for the identical particles case as follows

\begin{figure}\label{f:identicas}
\begin{center}
\includegraphics[width=7.5cm]{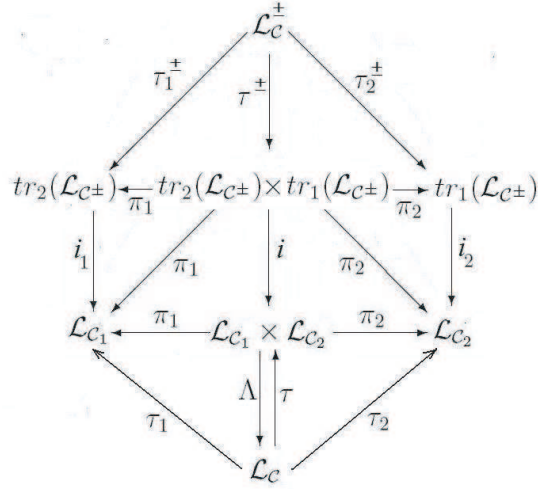}
\caption{\small{Canonical maps between the lattices
$\mathcal{L}_{\mathcal{C}}$, $\mathcal{L}_{\mathcal{C}_1}$,
$\mathcal{L}_{\mathcal{C}_2}$ and $\mathcal{L}_{\mathcal{C}^{\pm}}$.
The arrows $i$ and $\pi$ represent inclusions and canonical
projections.}}
\end{center}
\end{figure}

\begin{eqnarray}\label{e:Tauimasmenos}
\tau_{i}^{\pm}:\mathcal{L}_{\mathcal{C}^{\pm}}\longrightarrow
\mbox{tr}_j(\mathcal{L}_{\mathcal{C}^{\pm}})\nonumber\\
C\longmapsto \mbox{tr}_{j}(C)
\end{eqnarray}

\noindent and the product map

\begin{eqnarray}\label{e:Taumasmenos}
\tau^{\pm}:\mathcal{L}_{\mathcal{C}^{\pm}}\longrightarrow
\mbox{tr}_2(\mathcal{L}_{\mathcal{C}^{\pm}})\times
\mbox{tr}_1(\mathcal{L}_{\mathcal{C}^{\pm}})\nonumber\\
C\longmapsto(\tau_{1}^{\pm}(C),\tau_{2}^{\pm}(C))
\end{eqnarray}

\noindent In other words, the maps
$\tau_{i}^{\pm}=\tau_{i}|_{\mathcal{L}_{\mathcal{C}^{\pm}}}$ and
$\tau^{\pm}=\tau|_{\mathcal{L}_{\mathcal{C}^{\pm}}}$ are the
restrictions of $\tau_{i}$ and $\tau$ respectively to
$\mathcal{L}_{\mathcal{C}^{\pm}}$. Given that the subsystems are
identical, it follows that $\tau_{1}^{\pm}=\tau_{2}^{\pm}$. The map
$\tau^{\pm}$ is defined in analogy with \eqref{e:Tau}, and it allows
to link states of the compound system to states of its subsystems.

\noindent It would be of great interest to find an extension
$\Lambda^{\pm}$ of the canonic map $\Lambda$. However, such an
extension for the identical particle case is not immediate, because
the elements of the set $\mbox{tr}_2(C_{1})\otimes
\mbox{tr}_1(C_{2})$ will not be symmetrized states in the general
case.

\noindent The connection between the lattices
$\mathcal{L}_{\mathcal{C}_1}$, $\mathcal{L}_{\mathcal{C}_2}$ and
$\mathcal{L}_{\mathcal{C}}$ with the lattice of the identical
particles $\mathcal{L}_{\mathcal{C}^{\pm}}$ is shown in Figure
\ref{f:identicas}.

\section{Quasi-Set Theory}\label{s:Quasi-sets}

In this section, we basically follow the exposition of
\cite{kraare10}; for details, see \cite{frekra11},
\cite[Chap.7]{French-Kra-2006}. Quasi-set theory is a mathematical
theory that enables us to deal with collections (quasi-sets) of
objects that may be indiscernible without turning to be identical
(being the same object). The only way of dealing with objects of
this kind in a standard theory such as ZFC is to allow  the
introduction of some \ita{ad hoc} devices such as the restriction of
the lexicon of  properties, that is, by taking a language with a
finite number of them. This is Quine's famous way of defining
identity, namely, by the exhaustion of the chosen (finitely many)
predicates \cite[Chap.12]{qui82}. But this just defines
indiscernibility relative to the language's predicates, and not
identity strictly speaking, for there may exist other predicates not
in the language which distinguish among the entities.

In ZFC (the same can be said of most theories with due
qualification) given an object $a$, we can always form the unitary
set $A = \{a\}$ and define  a unary property (a formula with just
one free variable) of $a$ by posing $I_a(x)$ iff $x \in A$. This
formula of course distinguishes (or discriminates) $a$ from any
\textit{other} object for only $a$ satisfies it. Saying in other
words, in the standard mathematics (in the sense mentioned in the
Introduction), whenever we have a set with cardinal $\alpha > 1$,
its elements \textit{are} distinct. In short, there are no
indiscernible objects, except with respect to a few chosen
predicates.

Quine says that objects may be (strongly) discriminable when there
is a formula with only one free variable that is satisfied by one of
the objects but not by the other. Two objects are \ita{moderately
discriminable} when there exists a formula with two free variables
that is satisfied by the two objects in one order but not in the
other order \cite[Chap.15]{qui82}. As he recalls, any two real
numbers are strongly discriminable, although they may be
`specified', that is, definable by a formula. But in the
interpretation which supports the received view, there is no way to
attribute an identity to quanta (let us call this a `which is which
criterion'), something that isolate one of them from the others in
such a way that \ita{this} chosen object remains with its identity
forever.

Thus, how can we deal with, say, the two electrons of an Helium atom
in the fundamental state? We know that they have all the same
properties but distinct spins in a given direction, and so they obey
an irreflexive and symmetric relation `to have different spin of'
and thus they are moderately discriminable in the sense of Quine.
Muller and Saunders think that with this example, they have found
objects (the mentioned electrons) that are moderately but not
strongly discriminable \cite{mulsau08}. Thus apparently we have
found a way of speaking about two electrons with different spins and
cannot say which is which. But this is a false supposition (see
\cite{kra10}). Within classical logic (and the mentioned authors
assume that they are using $ZFC$), being objects (in some sense of
the word) in a finite number, they can always be named, say $a$ and
$b$ and the above property `to belong to $A = \{a\}$', which is true
just for $a$, distinguish them absolutely! In Quine's words, they
can be always specified, that is, the language (of $ZFC$) there is
always a formula in one free variable that is uniquely satisfied by
a given object \cite[p.134]{qui82}. You may say that this property
$I_a(x)$ (`being identical with $a$') is not a  `legitimate
property' of $x$, but in our opinion this would be a quite arbitrary
answer: which would be the legitimate ones? And why such an election
(if there is one)?

Quasi-set theory offers a different alternative to treat these
questions. Although electrons do present a difference due to Pauli's
exclusion principle (for instance, differences in their spins), any
permutation of them does not change the relevant probabilities, as
is well known; in other words, there is no a which is which
criterion. So, it seems  that it is better (and apparently most
correct) to say that the electrons may be indiscernible but without
assuming that this makes them the very same object. A paradigmatic
example may be that of the atoms in a Bose-Einstein condensate.
Thus, we need to avoid that indiscernibility implies identity (in
the philosophical sense of being the same entity). To cope with this
idea, we `separate' the concepts: indiscernibility, or
indistinguishability, is a relation that holds for all objects of
our domain, but identity is not. Certain objects may be
indiscernible without turning to be the same object, as implied by
the standard theory of identity, and they may form collections with
cardinals greater than one (this is achieved by the postulates of
the theory) but in a way that they cannot be identified, named,
labeled, counted in the standard way.\footnote{We mean: a series of,
say, five objects can be counted by proving that the set having them
as elements (and no other element) is equinumerous to a finite
ordinal, in the case, the ordinal $5 = \{0, \ldots, 4\}$. We remark
that in order to define the bijection, we need to distinguish among
the elements being counted --they need to be \ita{individuals}, yet
sometimes not specified.}  There is no space here to provide the
details of the theory, and we suggest \cite{French-Kra-2006} and
\cite{frekra11} for detailed references and for the axioms. Anyhow,
in the next subsection we outline without the details the main ideas
of the theory.

\subsection{Basic Ideas Of The Theory $\mathfrak{Q}$}

Intuitively speaking, a quasi-set is a collection of objects such
that some of them may be indistinguishable without turning to be
identical.

As mentioned above, quasi-set theory \Q\ has its main motivations in
some insights advanced by Schr\"odinger in that the concept of
identity would make no sense when applied to elementary particles
\cite[pp.\ 17-18]{sch52}. Another motivation is (in our opinion) the
need, stemming from philosophical worries, of dealing with
collections of absolutely indistinguishable items that  need not be
the same ones. Spatio-temporal differences could be used in this
case, you may say, and this serves do distinguish them. Without
discussing the role of spatio-temporal properties here (but see
\cite{French-Kra-2006}), we can argue that these objects are
invariant by permutations; in other words, the world does not
present differences in substituting them one from the other. Thus,
it would be difficult to say that they have some form of identity,
for they (in principle) lack any identifying characteristic. Objects
of this kind act like those that obey Bose-Einstein statistics, that
is, bosons (we should remember that the indiscernibility hypothesis
was essential in the derivation of Planck's formula --see
\cite{French-Kra-2006} once more).

Thus, the first point is to guarantee that identity and
indistinguishability (or indiscernibility) will not collapse into
one another when the theory is formally developed. We assume that
identity (symbolized by `=') is not a primitive relation, but we use
a weaker primitive concept of indistinguishability (symbolized by
`$\equiv$') instead. This is just an equivalence relation and holds
among all objects of the considered domain. If the objects of the
theory are divided up into groups, namely, the $m$-objects (standing
for `micro-objects') and $M$-objects (for `macro-objects')
---these are ur-elements--- and quasi-sets of them (an probably
having other quasi-sets as elements as well), then identity (having
all the properties of standard identity of ZF) can be defined for
$M$-objects and quasi-sets having no $m$-objects in their transitive
closure (this concept is like the standard one). Thus, if we take
just the part of the theory obtained by ruling out the $m$-objects
and collections (quasi-sets) having them in their transitive
closure, we get a copy of ZFU (ZF with \ita{Urelemente}); if we
further eliminate the $M$-objects, we get a copy of the `pure' ZF.

Technically, expressions such as $x=y$ are not always well formed,
for they are not formulas when either $x$ or $y$ denote $m$-objects
(entities satisfying the unary primitive predicate $m$). We express
that by informally saying that the concept of identity \ita{does not
always make sense} for all objects (it should be emphasized that
this is just a way of speech). The objects (the $m$-objects) to
which the defined concept of identity does not apply are termed
\ita{non-individuals} for historical reasons (see
\cite{French-Kra-2006}). As a result (from the axioms of the
theory), we can form collections of $m$-objects which have no
identity; these collections may have a cardinal (termed its
`quasi-cardinal') but not an associated ordinal. Thus, the concept
of ordinal and of cardinal are taken as independent, as in some
formulations of ZF proper. So, informally speaking, a quasi-set of
$m$-objects is such that its elements cannot be identified by names,
counted, ordered, although there is a sense in saying that these
collections have a cardinal (that cannot be defined by means of
ordinals, as usual, which presupposes identity).

When \Q\ is used in connection with quantum physics, the $m$-objects
are thought of as representing quanta (henceforth, \ita{q-objects}),
but they are not necessarily `particles' in the standard
sense---associated with classical physics of even with orthodox
quantum mechanics; waves, field excitations (the `particles' in
quantum field theory---QFT), perhaps even strings or whatever
entities supposed indiscernible can be taken as possible
interpretations of the $m$-objects. Generally speaking, whatever
`objects' sharing the property of being indistinguishable  can also
be values of the variables of \Q\ (see \cite[Chap.\ 6]{fal07} for a
survey on the various different meanings that the word `particle'
has acquired in connection to quantum physics).

Another important feature of \Q\ is that standard mathematics can be
developed using its resources, for  the theory is conceived in such
a way that ZFU (and hence also ZF, perhaps with the axiom of choice,
ZFC) is a subtheory of \Q. In other words, the theory is constructed
so that it extends standard Zermelo-Fraenkel with
\textit{Urelemente} (ZFU); thus standard sets (of ZFU) can be viewed
as particular qsets (that is, there are qsets that have all the
properties of the sets of ZFU, while the objects in \Q\
corresponding to the \textit{Urelemente} of ZFU are termed
$M$-atoms; these satisfy another primitive unary predicate $M$). The
`sets' in \Q\   will be called \Q-sets, or just \emph{sets} for
short. An object is a qset when it is neither an $m$-object nor an
$M$-object and, to make the distinction, the language of \Q\
encompasses a third unary predicate $Z$ such that $Z(x)$ says that
$x$ is a qset which is also a \ita{set}, and they correspond to
those objects erected in the `classical' part of the theory (without
$m$-objects). It is also possible to show that there is  a
translation from the language of ZFU into the language of \Q\, so
that the translations of the postulates of ZFU turn to be theorems
of \Q; thus, there is a `copy' of ZFU in \Q, and we refer to it as
the `classical' part of \Q. In this copy, all the usual mathematical
concepts can be stated, as for instance, the concept of ordinal (for
sets).

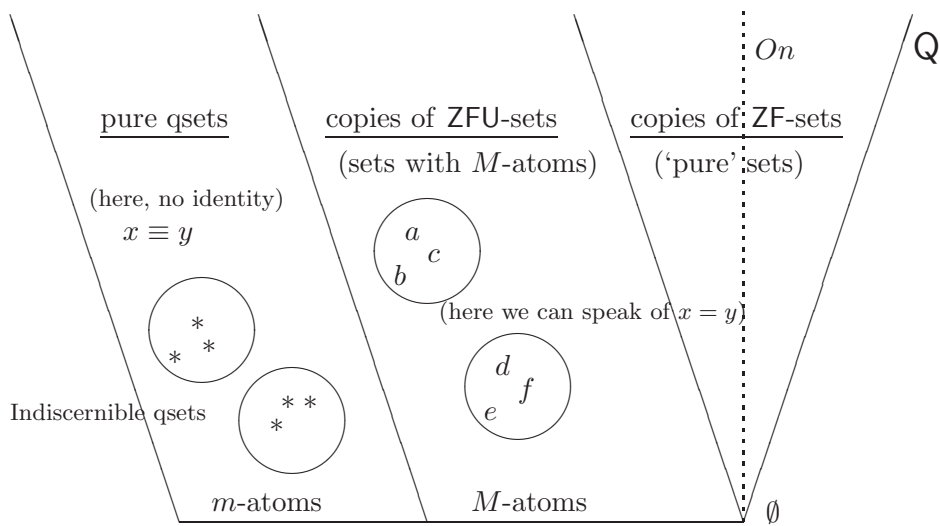
\begin{figure}[h]
\setlength{\unitlength}{1.5mm} \centering
\begin{picture}(80,60)

\put(15,5){\line(-1,3){15}}

\put(15,5){\line(1,0){50}}

\put(65,5){\line(1,3){15}}

\put(65,5){\line(-1,3){15}}

\put(37,5){\line(-1,3){15}}

%\put(37,5){\line(1,0){28}}

\put(80,46){\Large{$\mathsf{Q}$}}

\put(17,6){{ $m$-atoms}}

\put(40,6){{ $M$-atoms}}

\put(8,40){\underline{pure qsets}}

\put(10,30){$x \equiv y$}

\put(17,22){\circle{15}}

\put(7,33){\footnotesize{(here, no identity)}}
%------------------------
\put(17,20){$\ast$}

\put(16,22){$\ast$}

\put(14,19){$\ast$}
%--------------------------
\put(25,14){\circle{15}}

\put(23,13){$\ast$}

\put(26,15){$\ast$}

\put(24,15){$\ast$}

\put(0,14){\footnotesize{Indiscernible qsets}}
%---------------------------
\put(37,29){\circle{15}}

\put(35,30){$a$}

\put(34,26){$b$}

\put(37,28){$c$}

%---------------------------
\put(45,17){\circle{15}}

\put(43,18){$d$}

\put(42,14){$e$}

\put(45,16){$f$}

%------------------------------
\put(67,5){$\emptyset$}

\multiput(65,5)(0,1){46}{\bf\line(0,1){.3}}

\put(66,46){$On$}

\put(55,40){{\underline{copies of \textsf{ZF}-sets}}}

\put(57,36){(`pure' sets)}

\put(28,40){{\underline{copies of \textsf{ZFU}-sets}}}

\put(29,36){(sets with $M$-atoms)}

\put(38,23){\footnotesize{(here we can speak of $x=y$)}}

\end{picture}

\caption{The Quasi-Set Universe $\mathsf{Q}$: $On$ is the class of ordinals,
defined in the classical part of the theory. See \cite{kraare10}}\label{qsetuniverse}
\end{figure}

Furthermore, it should be recalled that the postulates for the
relation of indiscernibility, when applied to $M$-atoms or to
\Q-sets, collapses into standard identity (of ZFU). The \Q-sets are
qsets whose transitive closure (defined as usual) does not contain
$m$-atoms (in other words, they are `constructed in the classical
part of the theory---see Fig.\ \ref{qsetuniverse}).

%-------------------------------

In order to distinguish between \Q-sets and qsets that may have
$m$-atoms in their transitive closure, we write (in the
metalanguage) $\{x : \varphi(x)\}$ for the former and $[x :
\varphi(x)]$ for the latter. In \Q, we term `pure' those qsets that
have only $m$-objects as elements (although these elements may be
not always indistinguishable from one another, that is, the theory
is consistent with the assumption of the existence of different
kinds of  $m$-atoms), and to them it is assumed that the usual
notion of identity cannot be applied (that is, let us recall, $x=y$,
as well as its negation, $x \not= y$, are not well formed formulas
if either $x$ or $y$ stand for $m$-objects). Notwithstanding, the
primitive relation $\equiv$ applies to them, and it has the
properties of an equivalence relation.

We have also a defined concept  of \textit{extensional identity} and
it has the properties of standard identity of ZFU. More precisely,
we write $x =_E y$ (read '$x$ and $y$ are extensionally identical')
iff they are both qsets having the same elements (that is, $\forall
z (z \in x \lra z \in y)$) or they are both $M$-atoms and belong to
the same qsets (that is, $\forall z (x \in z \lra y \in z)$). From
now on, we shall not bother to always write $=_E$, using simply the
symbol ``='' for the extensional equality.

Since $m$-atoms cannot be identified in the formalism, it is not
possible in general to attribute an ordinal to qsets of such
elements. Thus, for certain qsets, it is not possible to define a
notion of cardinal number by means of ordinals. The theory uses  a
primitive concept of \ita{quasi-cardinal} instead, which intuitively
stands for the `quantity' of objects in a collection.\footnote{A
notion of finite quasi-cardinal can be defined as a derived concept
(see \cite{Dom-Hol-2007}).} The theory has still an `axiom of weak
extensionality', which states (informally speaking) that those
quasi-sets that have the same quantity of elements of the same sort
(in the sense that they belong to the same equivalence class of
indistinguishable objects) are indistinguishable by their own. One
of the interesting consequences of this axiom is  related to the non
observability of permutations in quantum physics, which is one of
the most basic facts regarding indistinguishable quanta (for a
discussion on this point, see \cite{frerick03}). In standard set
theories, if $w \in x$, then of course $(x - \{w \}) \cup \{z\} = x$
iff $z = w$. That is, we can 'exchange' (without modifying the
original arrangement) two elements iff they are \textit{the same}
elements, by force of the axiom of extensionality. In \Q\ we can
prove the following theorem, where $[[z]]$ (and similarly $[[w]]$)
stand for a quasi-set with quasi-cardinal 1 whose only element is
indistinguishable from $z$ (respectively, from $w$ --the reader
shouldn't think that this element \textit{is identical to either}
$z$ or $w$:

\begin{theorem}[Unobservability of Permutations]\label{unobservabilty}
 Let $x$ be a finite quasi-set such that $x$ does not
contain all indistinguishable from $z$, where $z$ is an $m$-atom
such that $z \in x$. If $w \equiv z$ and $w \notin x$, then there
exists $[[w]]$ such that
$$(x - [[z]]) \cup [[w]] \equiv x$$
\end{theorem}

\noindent Informally speaking, supposing that $x$ has $n$ elements,
then if we `exchange' their elements $z$ by corresponding
indistinguishable elements $w$ (set theoretically, this means
performing the operation $(x - [[z]]) \cup [[w]]$), then the
resulting quasi-set remains \textit{indistinguishable} from the one
we started with. In a certain sense, it does not matter whether we
are dealing with $x$ or with $(x - [[z]]) \cup w'$.  So, within \Q,
we can express that `permutations are not observable', without
necessarily introducing symmetry postulates, and in particular to
derive `in a natural way' the quantum statistics (see
\cite[Chap.7]{French-Kra-2006}).

\subsection{The $\mathfrak{Q}$-space}\label{s:Q-space}

As we have seen in Section \ref{s:IdenticalParticles}, if particles
are indistinguishable they can only access symmetrized states. But
particles are labeled in this construction, and this procedure was
criticized (Section \ref{s:PosingTheProblem}). It has been claimed
that the Fock-space formalism poses a solution to the questions
raised by this criticism \cite{redtel91,redtel92}. But the
Fock-space formalism also makes use of particle labeling uses
particle labeling in order to obtain the correct states
\cite{French-Kra-2006}.

How can we avoid this problem of the Fock-space formulation of $QM$?
If we could avoid the individuation of the particles at every step
of the construction of a Fock-like formulation of $QM$, we would
give a positive answer to the problem posed in
\cite{redtel91,redtel92} (recalled in the Introduction) which is not
affected by the criticisms linked to it. Quasi-set theory can be
used for this purpose, and in fact, this construction has been done
\cite{Dom-Hol-Kra-2008,Dom-Hol-Kniz-Kra-2009}. There, an alternative
proposal is presented which resembles that of the Fock-space
formalism but based on $\mathfrak{Q}$. And thus, genuinely avoiding
artificial labeling. We give a sketch of the construction here,
mainly following \cite{Dom-Hol-Kra-2008,Dom-Hol-Kniz-Kra-2009}. And
in the following Section, we will also see that this kind of
constructions not only allows us to solve the problem posed above,
but they serve also to plant interesting foundational issues.

Let us consider a set $\epsilon= \{\epsilon_{i}\}_{i \in I }$ (that
is, a ``set" in $\mathfrak{Q}$), where $I$ is an arbitrary
collection of indexes (this makes sense in the `classical part' of
$\mathfrak{Q}$). Suppose that the elements $\epsilon_{i}$ to
represent the eigenvalues of a physical observable. Next,
quasi-functions $f$ are constructed, such that $f:\epsilon
\longrightarrow \mathcal{F}_{p}$, where $\mathcal{F}_{p}$ is the
quasi-set formed of finite and pure quasi-sets. $f$ is a quasi-set
formed by ordered pairs $\langle \epsilon_{i};x\rangle$ with
$\epsilon_{i}\in\epsilon$ and $x\in\mathcal{F}_{p}$.

These quasi-functions are chosen in such a way that whenever
$\langle \epsilon_{i_{k}};x\rangle$ and $\langle
\epsilon_{i_{k'}};y\rangle$ belong to $f$ and $k\neq  k'$, then
$x\cap y=\emptyset$. It is further assumed that the sum of the
quasi-cardinals of the quasi-sets which appear in the image of each
of these quasi-functions is finite. This means that $qc(x)=0$ for
every $x$ in the image of $f$, except for a finite number of
elements of $\epsilon$. These quasi-functions form a quasi-set
called $\mathcal{F}$.

A pair $\langle \epsilon_{i};x\rangle$ is interpreted as the
statement ``the energy level $\epsilon_{i}$ has occupation number
$qc(x)$". Quasi-functions of this kind are represented by
expressions such as
$f_{\epsilon_{i_{1}}\epsilon_{i_{2}}\ldots\epsilon_{i_{m}}}$. If the
symbol $\epsilon_{i_{k}}$ appears $j$-times the level
$\epsilon_{i_{k}}$ has occupation number $j$. The levels that do not
appear have occupation number zero.

At this point of the construction, the indexes appearing in
$f_{\epsilon_{i_{1}}\epsilon_{i_{2}}\ldots\epsilon_{i_{n}}}$ has no
meaning at all. But an order can be defined as follows. Given a
quasi-function $f\in\mathcal{F}$, let
$\{\epsilon_{i_{1}}\epsilon_{i_{2}}\ldots\epsilon_{i_{m}}\}$ be the
quasi-set formed by the elements of $\epsilon$ such that
$\langle\epsilon_{i_{k}},x\rangle\in f$ and $qc(x)\neq  0$ ($k=
1\ldots m$). This quasi-set is denoted $supp(f)$. Consider now the
pair $\langle o,f\rangle$, where $o$ is a bijective quasi-function
$o:\{\epsilon_{i_{1}}\epsilon_{i_{2}}\ldots\epsilon_{i_{m}}\}\longrightarrow
\{1,2,\ldots,m\}$. Each one of the quasi-functions $o$ defines an
order on $supp(f)$. Let $\mathcal{O}\mathcal{F}$ denote the
quasi-set formed by all the pairs $\langle o,f\rangle$.
$\mathcal{O}\mathcal{F}$ is the quasi-set formed by all the
quasi-functions of $\mathcal{F}$ with ordered support. Using a
similar notation as before (and also repeating indexes according to
the occupation number),
$f_{\epsilon_{i_{1}}\epsilon_{i_{2}}\ldots\epsilon_{i_{n}}}\in
\mathcal{O}\mathcal{F}$ refers to a quasi-function $f\in\mathcal{F}$
and a special ordering of
$\{{\epsilon_{i_{1}}\epsilon_{i_{2}}\ldots\epsilon_{i_{n}}}\}$. But
now, the order of the indexes must not be understood as a labeling
of particles, because it can be shown that the permutation of
particles does not give place to a new element of
$\mathcal{O}\mathcal{F}$ \cite{Dom-Hol-Kra-2008}.

Consider next the collection of quasi-functions $C$ which assign to
every $f\in \mathcal{F}$ (or $f\in \mathcal{O}\mathcal{F}$) a
complex number. A quasi-function $c\in C$ is a collection of ordered
pairs $\langle f;\lambda\rangle$, where $f\in \mathcal{F}$ (or $f\in
\mathcal{O}\mathcal{F}$) and $\lambda$ a complex number. Let $C_{0}$
be the subset of $C$ such that, if $c\in C_0$, then $c(f)=0$ for
almost every $f\in \mathcal{O}\mathcal{F}$ (i.e., $c(f)=0$ for every
$f\in \mathcal{O}\mathcal{F}$ except for a finite number of
quasi-functions). A sum and a product can be defined in $C_{0}$ as
follows

\begin{definition}
Given $\alpha,\beta,\gamma\in \mathcal{C}$, and $c,c_{1},c_{2}\in
C_{0}$, then
$$(\gamma\ast c)(f) := \gamma(c(f))\ \ and \ \
(c_{1}+c_{2})(f) :=  c_{1}(f) + c_{2}(f)$$
\end{definition}

\noindent Using the above definitions, $(C_{0},+,\ast)$ is endowed
with a complex vector space structure. Given a quasi-function $c\in
C_{0}$ such that $c(f_{i})= \lambda_{i}$ ($i= 1,\ldots,n$) for some
finite set of quasi-functions $\{f_{i}\}$ belonging to $\mathcal{F}$
or $\mathcal{O}\mathcal{F}$ the following association is done

\begin{equation}
c\approx(\lambda_{1}f_{1}+\lambda_{2}f_{2}+\cdots+\lambda_{n}f_{n})
\end{equation}

\noindent Thus, a quasi-function $c\in C_0$ is interpreted as a
linear combination of the quasi-functions $f_{i}$ (representing a
quantum superposition).

Scalar products must be introduced in order to reproduce the quantum
mechanical machinery of computation of probabilities. It is possible
to define two of them, one for bosons (``$\circ$") and one for
fermions (``$\bullet$"). In this way (and using norm completion),
two Hilbert spaces $({\mathbb{V}_{Q}},\ \circ )$ and
$({\mathbb{V}_{Q}},\ \bullet )$ are obtained. The scalar product for
bosons is defined as follows

\begin{definition}
Let $\delta_{ij}$ be the Kronecker symbol and
$f_{\epsilon_{i_{1}}\epsilon_{i_{2}}\ldots\epsilon_{i_{n}}}$ and
$f_{\epsilon_{i'_{1}}\epsilon_{i'_{2}}\ldots\epsilon_{i'_{m}}}$ two
basis vectors, then
$$
f_{\epsilon_{i_{1}}\epsilon_{i_{2}}\ldots\epsilon_{i_{n}}}\circ
f_{\epsilon_{i'_{1}}\epsilon_{i'_{2}}\ldots\epsilon_{i'_{m}}} :=
\delta_{nm}\sum_{p}\delta_{i_{1}pi'_{1}}\delta_{i_{2}pi'_{2}}\ldots\delta_{i_{n}pi'_{n}}
$$
The sum is extended over all the permutations of the set
$i'=(i'_{1},i'_{2},\ldots,i'_{n})$ and for each permutation $p$,
$pi'=(pi'_{1},pi'_{2},\ldots,pi'_{n})$.
\end{definition}

\noindent and for fermions

\begin{definition}
Let $\delta_{ij}$ be the Kronecker symbol,
$f_{\epsilon_{i_{1}}\epsilon_{i_{2}}\ldots\epsilon_{i_{n}}}$ and
$f_{\epsilon_{i'_{1}}\epsilon_{i'_{2}}\ldots\epsilon_{i'_{m}}}$ two
basis vectors, then
$$f_{\epsilon_{i_{1}}\epsilon_{i_{2}}\ldots\epsilon_{i_{n}}}\bullet
f_{\epsilon_{i'_{1}}\epsilon_{i'_{2}}\ldots\epsilon_{i'_{m}}} :=
\delta_{nm}\sum_{p}s^{p}\delta_{i_{1}pi'_{1}}\delta_{i_{2}pi'_{2}}\ldots\delta_{i_{n}pi'_{n}}$$
where: $s^{p}=+1$ if $p$ is even and $s^{p}= -1$ if $p$ is odd.
\end{definition}

\noindent These products can be easily extended to all linear
combinations. The second product $\bullet$ is an antisymmetric sum
of the indexes which appear in the quasi-functions and the
quasi-functions must belong to $\mathcal{O}\mathcal{F}$. If the
occupation number of a product is greater or equal than two, then,
it can be shown that the vector has null norm. Thus reproducing
Pauli's exclusion principle for fermions \cite{Dom-Hol-Kra-2008}.

With these constructions within $\mathfrak{Q}$, the formalism of
\textit{QM} can be rewritten giving a positive answer to the problem
of giving a formulation of \textit{QM} in which intrinsical
indistinguishability is taken into account from the beginning,
without artificially introducing artificial labels
\cite{Dom-Hol-Kniz-Kra-2009,Dom-Hol-Kniz-Kra-2009}.

\section{Stating the problem in an adequate
form}\label{s:WhatIsToBeDone?}

Once that the formal setting is determined, we are now ready to come
back to the questions posed in Section \ref{s:PosingTheProblem} in a
more formal way. In this Section we state the problem of identical
particles from a new perspective.

\subsection{Metaphysical undetermination}

As is well known, there are several interpretations of \textit{QM},
the received view being only one among others (perhaps, the most
popular). The simple fact that there exists an interpretation such
as Bohm's --in which particles possess definite trajectories--
represents a problem for someone who wants to extract metaphysical
constructions out of physical theories: how to reconcile
incompatible but plausible interpretations of a given formalism
which reproduces laboratory experiences, such as the von Neumann
formulation of \textit{QM}? While in the Bohmian interpretation
particles have definite trajectories, there are no trajectories at
all in the standard interpretation. Particles are individuals for
Bohm and non-individuals for the received view. This is the problem
of metaphysical underdetermination, discussed in detail in reference
\cite{French-Kra-2006}. We will review this problem here.

While Schr\"{o}dinger used the BE and FD statistics as an argument
to support the received view, other authors, interpreted these
``strange" statistics as a new form of non-local correlation between
particles (considered as individuals). While the symmetrization
postulate was used as an argument against particle individuality,
Muller and Saunders use particle labels to show that quanta are
individuals \cite{MullerSaubders-2008,MullerSeevinck-2009}. There
seems to be reasonable arguments for both positions: both
interpretations, while incompatible, seem to be valid, in the sense
that they do not contradict (up to now) empirical data. The received
view has historical problems such as stating clearly what does a
non-individual mean. But there are concrete solutions to this
problem, see for example \cite{French-Kra-2006}.

In addition to this \emph{metaphysical} undetermination, in Section
\ref{s:Q-space} we showed that an alternative formulation of
\textit{QM} may be given, but with a different underlying logic,
based on quasi-set theory. This implies that there also exists a
kind of \emph{logical} underdetermination, i.e., there is no
preferred logic, if the objective is to formulate the theory in an
axiomatic way.

Regarding \emph{logical underdetermination} it is important to
specify what we mean by the word ``logical". As is well known,
language has different layers. The axioms of a physical theory,
stated in a mathematical form, have an underlying logic, which is
usually standard set theory, but it may be formulated in a different
frame, such as category theory or even higher order logic. But here
we shall speak of ZFC set theory only. Then, the word ``logic" means
the axioms used in the mathematical formulation of the theory. $ZFC$
set theory has a deeper logical level, which are the axioms of first
order classical logic.

But physics not only concern mathematical formulation.
Interpretation may be regarded as part of the theory or not, but it
is for sure that it is unavoidable to have, at least, a minimal
interpretational framework in order to connect theory with
experience (and eventually, to explain it). The language used for
speaking (and thinking) about the concepts related to the word
``experience", as well as the theoretical terms representing the
entities involved in a given interpretation, are not just
mathematical, and it is also not an artificial one based on first
order classical logic. But this language has its own ``underlying
logic", which may be not necessarily a formalized one. What does
Scrh\"{o}dinger meant by a non-individual entity, may be not clear
or formal, but it is clear that the logic underlying such an
interpretation seems to be not the classical one. Quasi-set theory,
would provide a formal framework in order to give us a formal
logical stuff for that notion. The logic at the level of the axioms
of the theory and the underlying logic of the interpretation, may
coincide or not. The second one is the case of the received view,
susceptible of all the criticisms mentioned above. $QM$ formulated
in the $\mathfrak{Q}$-space formalism seems to be in harmony with
the underlying logic of the received view.

We are thus faced with incompatible alternatives to take, with
different possibilities. How to make a choice? Which attitude is to
be taken in the view of this fact?

\subsection{Ockham's razor revisited}

In order to answer the question posed in the previous Section, let
us review a traditional example of quantum theory. Bohm's
interpretation presupposes individual trajectories for quantum
particles, guided by pilot waves. Quantum statistics would thus be
ruled out by hidden variables: trajectories exist, but you will
never be able to predict them. The same interpretation comes endowed
with a ``concealment mechanism", which forbids experimental control
of the postulated hidden variables. Thus, these hidden variables,
while compatible with predicted experience, play no role in any
experiment, i.e., they are completely \emph{dispensable} (for the
received view). They play only an explanatory role in Bohm´s
approach: fairies or elves may be responsible for the values that
these hidden variables take. \emph{But the impossibility of settling
this question experimentally is an a priori requirement of the
interpretation itself}.

In view of the discussion of the previous Section, the word
``explanatory", should be understood as follows: to make experience
compatible with an ontological (or metaphysical) ``preference" (or
``prejudice"). Any interpretation seems to have theoretical terms
which may be suppressed in order to endorse a minimal
interpretation. But this is not the point that we want to
stress:\emph{ dispensability is not our problem}. What we want to
remark is that hidden variables in Bohm's theory cannot be measured,
nor controlled in any laboratory experience. This is precisely an
unavoidable requirement of the Bohmian interpretation, in order to
be formally equivalent to standard \emph{QM}. Otherwise, if these
variables could be controlled, or some crucial experiment based on
them could be designed, standard \emph{QM} would be wrong (and the
supposed formal equivalence between Bohm´s theory an the orthodox
formulation of QM would no longer hold). Alike a quantum state or
the electron charge, or even a field --which are all theoretical
terms--, hidden variables cannot be prepared nor measured, simply
because this is what ``hidden" is intended to mean.

Which is the limit? There is no limit. Given a metaphysical
preference (or election), we can always add as many ``hidden"
theoretical terms as we want, always taking care that they should
not make predictions incompatible with experience (which is
regulated by formalism plus a minimal interpretation). But it seems
reasonable to assume that science should not be concerned with
notions which are
---as a matter of principle--- impossible to control in any
experiment, adding neither new predictions nor postulating states of
affairs which are \emph{by definition} impossible to regulate
experimentally. It could be the case that a notion used to explain
phenomena, such as Boltzman´s particles, could not be observed or
clearly studied in any experimental set up at a certain stage of a
theory, \emph{but this impossibility cannot be part of the
definition of this notion}.

As is well known, the existence of Bohm's interpretation and the
fact that its hidden variables are non-local, led J. Bell to
question himself if there may exist an interpretation based on local
hidden variables. These questionings originated the well known story
about Bell's inequalities and Bell's theorem: any interpretation
based on hidden variables compatible with quantum predictions is
attained to non-locality, i.e., \emph{hidden variables must be
non-local}. And this was tested experimentally (in favor of $QM$).
Thus, Bell's theorem shows that the acceptance of hidden variables
leads also to ``hidden non-locality", which of course, cannot be
used to send information instantaneously.

One of the conclusions that we extract from the story of hidden
variables and Bell's theorem, is as follows. It is always possible
to make different interpretations of a given theory, and they may
result incompatible. Where does this metaphysical underdetermination
comes from? We will not discuss this in detail here, but we will
only stress one point. It may be possible that metaphysical
underdetermination of physical theories is a general characteristic
of language itself, which manifests even at the level of simple
examples of logic: think about models built within set theory. There
may be several models of the same axiomatic, and nothing determines
a preferred one. Another example, is one of the G\"odel's results,
which asserts that any axiomatic system ---with a certain degree of
complexity and formulated in a certain way--- has true but
undecidable propositions: there is always something beyond the scope
of the axioms. And if this happens already at the logical level,
nothing prevents this to happen in more complex languages, such as
the one employed in the formulation and interpretation of a physical
theory. The complete ``language" of a theory involves a complex
mixture of laboratory assertions, theoretical concepts (many of them
not necessarily completely or rigourously defined), and if we wish,
the formal language of the axiomatic level too.

\emph{Thus, it may well be that, as well as it happens with Godel's
theorem, there will always exist assertions which cannot be decided
using the axioms of the theory plus the available experimental data
at a given historical moment, and no one knows how to state the
problem in order to design the adequate experiments to decide them}.
But as it happened with hidden variables and Bell's theorem ---by
adding the adequate experimental evidence--- further non trivial
information about hidden variables was extracted by stating the
problem in an adequate form. Adequate formulation of the problem may
involve introducing new axioms and definitions (as well as
theoretical constructions), in order to distinguish between several
alternatives. This is one of the great merits of Bell.

A similar program may be delineated for indistinguishable particles.
Even if we do not know how to make an experimental test in order to
decide if quanta are individuals or not, working on formal
structures and by considering ontological specification of the
involved entities may serve to design new experiments, which if they
don't rule out a given possibility, they may impose restrictions on
its validity. As happened with Bohmmian mechanics: hidden variables
resulted to be non-local, and as this fact is in a certain sense
incompatible with restricted relativity theory, it gives us more
elements to make an election (thought we are no obliged to consider
it). A similar analysis can be made for the Kochen-Specker theorem.

Even now, we have at hands examples of how this ideas work for
identical particles. In order to explain BE and FD statistics, one
may assume non-individuality, as usual. But other authors explain
this phenomenon by postulating non-independent correlations (which
are different for Fermions and Bossons) \cite{vanFrassen}. In spite
of these small steps, we think that an analogue to Kochen-Speker or
Bell's theorem for non-individuality is in order. This should be
added to the program for the investigation of identical particles in
another work.

It is worthy to analyze what happens with individuality in $CM$.
Trajectories can be measured in $CM$, and thus, one may postulate
that any trajectory, it corresponds to one particle or to a body. By
means of this association, individuality plays a direct role in
observation. Of course, nothing prohibits us to break the link
between particles and trajectories and postulate that a strange
(unobservable) particle permutation may happen between the
trajectories, and then, individuality could be questioned. We could
go further and presuppose that classical particles of the same mass
and form, charge, etc., are completely indistinguishable, and that
they do not have individuality at all, but their trajectories do. In
this case, we force the interpretation in order to satisfy a certain
metaphysical preference, but none of these extra assumptions can be
manipulated experimentally. This example in $CM$ is the reverse of
that one in the quantum case: as well as Bohmian Hidden Variables,
individuality of quanta is not only dispensable, but it is also
impossible to manipulate experimentally.

A possible attitude towards this ``metaphysical freedom" could be:
use Ockham's razor and discard individuality of quanta. But there
exists an alternative and (we think) more fruitful possibility:
given extra assumptions (such as the individuality of quanta), we
must provide more precise definitions, add extra postulates and to
design adequate crucial experiments in order to discriminate and
discard between possible metaphysical alternatives.

Before entering into the general conclusions of this work, let us
make an interesting remark. Bohm's interpretation postulates
trajectories, hidden variables and pilot waves. At the end of the
story, when these assumptions are fully analyzed, we find that
hidden variables where non-local, and that the pilot waves suffer of
similar problems than that of the traditional Scr\"{o}dinger´s wave
functions. It is as if the ``problematic" aspects of the standard
interpretation of quantum mechanics reemerge in the Bohmian
interpretation in a new fashion, or as if there was a ``principle of
conservation of problematic aspects". Again, this kind of analysis
of the ``failure" of the Bohmian's program (i.e., the failure to
recover a completely classical picture), does not suffice to discard
the whole interpretation. But it sheds light to the consequences of
our metaphysical preferences, and by studying these consequences, we
gain a lot of information of how things really work. Perhaps the
interesting question is not ``which is the true ontology or
interpretation?"; but questions such as ``which are the consequences
of each of them?" and ``which of them are untenable (according to
experience) and which of them are compatible?". This is perhaps the
most interesting attitude towards the ``metaphysical freedom"
originated in metaphysical underdetermination.

\section{Final discussion}\label{s:OntologicalImplications}

As we have seen, non-classical logics and algebraic structures may
arise also from insights taken from science. Non-reflexive logics,
in the sense posed above, constitute a typical example. But, what
can we say about ontology? A traditional philosopher may guess that
this question is not well posed, for ontology is the study of the
basic stuff of the world and it would be indifferent to the theory
we are using in such an investigation. We may say that, in this old
sense, a philosopher would say that ontology is the study of the
basic furniture of the world, and in this sense all we need is to
put some light on this world's stuff. But, at least since Quine
\cite{Quine}, we become familiarized with the talk of an ontology
associated to a theory and in order to speak about ontology, we need
to look to our better theories and consider what they say about the
world. This \ita{naturalized ontology} is today well accepted by
most philosophers of science, and we believe that this way of speech
is not contrary with the usual assumptions made by the physicists.
Thus, taking into account quantum theory (either relativistic or
non-relativistic versions --- quantum field theories), we can have
some insights for instance about the very nature of the basic
constituents of the world, the `elementary particles'. By
`elementary particles' we mean whatever entity postulated or assumed
by the theory in its grounds. We can even name then: electrons,
protons, neutrinos, quarks, and so on, and refer to them
indiscriminately as \ita{quanta}.

As well as logical constructions may be inspired in physical
theories, if we adopt the above point of view about a
\emph{naturalized ontology}, we may use these logical constructions
to draw conclusions about the possible ontological commitments of
physical theories. Let us discuss next the implications of the
existence of the formal structures presented in this paper.

\subsection{$\mathfrak{Q}$ and non-individuality}

Regarding the discussion we have made in this paper, we can take
some facts from granted, as for instance:

\begin{enumerate}
\item Standard formulations or both non-relativistic and relativistic
quantum mechanics use both classical logic and standard mathematics (say, that one built in ZFC).
Hence, no theory founded on such a basis can contradict its theorems (that is, classical logic and standard mathematics).
\item Quanta of the same species may be (in certain situations) absolutely indiscernible by all means provided by the theory.
\item There is a sense in saying that, in certain situations, quanta cannot be said to have an
identity, a permanent label that distinguishes each one of them from others even of the similar species.
\end{enumerate}

\noindent From these three items, we can conclude some basic facts.
From (1), standard formalism of \textit{QM} always distinguishes
between two quanta, even if they are of the same species and
regarded as indistinguishable. If the distinction cannot be achieved
in the physical theory proper (the specific postulates of $QM$ we
are using), they can be discerned by the underlying mathematics. In
fact, two distinct things can be put within two disjoined open sets,
and this distinguishes them absolutely. Thus, even the atoms in a
Bose-Einstein condensate, when represented in the mathematical model
($QM$) constructed this way, can be distinguished. From the logical
point of view, we cannot agree \ita{in totum} with the Nobel Prize
winner Wolfgang Ketterle, when he says that

\begin{quote}
``[i]f we have a gas of ideal gas particles at high temperature, we
may imagine those particles to be billiard balls ($\ldots$). They
race around in the container and occasionally collide. This is a
classical picture. However, if we use the hypothesis of de Broglie
that particles are matter waves, then we have to think of particles
as wave packets. The size of a wave packet is approximately given by
the de Broglie wavelength $\lambda_{dB}$, which is related to the
thermal velocity $v$ of the particles as $\lambda_{dB} = h/mv$. Here
$m$ is the mass of the particles and $h$ is Plancks´s constant. Now,
as long as the temperature is high, the wavepacket is very small and
the concept of indistinguishability is irrelevant, because we can
still follow the trajectory of each wavepacket and use classical
concepts. However, a real crisis comes when the gas is cooled down:
the colder the gas, the lower the velocity, and the longer the de
Broglie wavelength. When individual wave packets overlap, then we
have an identity crisis, because we can no longer follow
trajectories and say which particle is which. At that point, quantum
indistinguishability becomes important and we need quantum
statistics.''\cite{Ket.07}
\end{quote}

\noindent In fact, within standard logic and mathematics, there is
no identity crisis! This leads us to (2) and (3), which may be taken
as emphasizing the same question, namely, the way to go around
standard logic (and mathematics) in order to acknowledge that there
is in fact an identity crisis involving quanta in certain
situations. As we have said above, there are two options: to confine
ourselves to a certain protected region within standard ZFC (say a
mathematical structure) and speak of \ita{some} properties and
relations only. Thus the structure may be nonrigid and we can define
indiscernible objects as those which are lead to one another by one
of the non-trivial automorphisms of the structure. This is the
classical solution: symmetric and anti-symmetric vectors do the job,
and assuming identity as defined \ita{a la} Quine completes the
crime \cite{Quine2}\footnote{ Quine defines identity by the
exhaustion of all predicates of the language, taken always in a
finite number. In our opinion, this strategy defines only
indiscernibility regarding the chosen predicates.}. But, let us
recall, any structure built in ZFC can be extended to a rigid
structure, that is, a structure whose only automorphism is the
identity function. In this structure, the very nature of our alleged
indiscernible quanta would be revealed, and in the background, we
can see them as individuals, as entities having identity.

Thus, we conclude that if we aim at to speak of an ontology of
really indiscernible quanta, we need to go out from classical
frameworks and adopt an alternative logic, and quasi-set theory is
one of the options. And in fact, as we have shown in section
\ref{s:Q-space}, the construction presented can be used to
reformulate $QM$ in a different logical background than that of
$ZFC$ (meaning standard mathematics and classical logic). An
important consequence of the existence of such a reformulation is
that the conclusions about the identity of quanta posed in (1) above
are not a characteristic of \emph{any} formulation of $QM$, but only
of those formulations based on $ZFC$ or similar ``classical''\
theories. The construction sketched above show us that the answer to
the problem posed in \cite{redtel91,redtel92} and recalled in the
Introduction is in the affirmative, and that our construction is
more in accordance with the interpretations of $QM$ which claim that
particles are not individuals. The formulation of $QM$ discussed in
section \ref{s:Q-space} \emph{supports} this ``received view'' (see
\cite{French-Kra-2006}).

It is not our aim in this article to deny hidden variable theories
(or haecceities). We only stress the point that there are different
interpretations of $QM$ and according to this fact, different
mathematical formulations, each of which is more or less compatible
with a given interpretation.

\subsection{Quantum logic for compound systems of identical particles}\label{s:QLconclussions}

The quantum logical approach to physics does not restrict itself to
$QM$. It is a general operational framework which allows us to
include a huge family of physical theories. Using the method
developed by Piron \cite{piron}, it is possible to define questions
on an arbitrary system, in such a way that these questions form
propositions. It is possible to show that these propositions form a
lattice. By imposing suitable axioms on this lattice, one may
recover $QM$, $CM$, or, in principle, any arbitrary theory. This was
called the Operational Quantum Logic approach to physics ($OQL$).
This kind of approach, allows us to study the structure of the
propositional lattices of any given system, in particular, it allows
us to study the structure formed by elementary tests in $QM$, which
as is well known since the work of Birkoff and von Neumann
\cite{BvN}, they are isomorphic to the projection lattice
$\mathcal{P}(\mathcal{H})$. At the same time, this structural
characterization allows for a comparison between theories: although
$CM$ and $QM$ are very different theories, the $OQL$ approach allows
to compare them in a same formal framework (that is, lattice
theoretical) in order to look for analogies and differences. The
most striking one is perhaps that the propositional lattice of $QM$
is no-distributive.

The difficulties appear when we realize that the $OQL$ approach has
problems when applied to compound systems. It is true that, from a
foundational point of view, it gives useful information about the
structure of compound systems, but it works in a negative way: this
is the content of the results of Aerts and others
\cite{aertsjmp84,aertsparadox,AertsTensorProduct,AertsTensor-84}.
The fact that no product of lattices exist, tell us a lot about the structure of
compound quantum systems, but this is because of the incapability of
the approach to describe them. This incapability comes from the fact
that when we have an entangled state of the compound system, the
reduced state will not be pure, and thus, there will be not possible
to link the state of the compound system to the states of the
subsystems at the level of lattices. This happens simply because the
(possibly) mixed states of the subsystems cannot be represented as
elements of the corresponding lattices
\cite{extendedql,Holik-Massri-Ciancaglini-2010,Holik-Massri-Plastino-Zuberman}.
And this is critically true for the case of indistinguishable
particles: if we depart from a pure symmetrized state of the
bipartite system of fermions, we will always obtain mixed states for
the subsystems. This is essentially the reason why the $OQL$ approach
presents some disadvantages for the study of entanglement and, even more, for
the case of identical particles.

The constructions presented in
\cite{extendedql,Holik-Massri-Ciancaglini-2010,Holik-Massri-Plastino-Zuberman}
overcome this difficulty, by incorporating mixed states as atoms of
a new lattice. This allows to link states of the compound system to
states of the subsystems, and show us the structure of the
propositions thus formed. Convex subsets of the convex set of states
may be interpreted as probability spaces
\cite{Holik-Massri-Plastino-Zuberman}: our constructions allow us to
look at the structure of these probability spaces. This was not
possible using the traditional $OQL$ approach.

In this work, we have shown that a quantum logical structure for
compound quantum systems of identical particles can be realized,
something which was not present in the literature excepting for
scarce examples \cite{Grigore,AertsIdentical}. Our structure
captures the maps which link the states of a compound system formed
of two identical particles to the states of its subsystems. Thus,
providing a formal framework in which we can study how compound
quantum systems of identical particles behave. In particular, they
allow us to see the divergences with classical structures. We hope
that these structures allow us to study the formal structure of
compound quantum systems of identical particles in future work.

\bigskip\bigskip
\noindent {\bf Acknowledgements} \noindent D. Krause is partially supported by
CNPq, grant 300122/2009-8.

\end{document}